\documentclass[lettersize,journal]{IEEEtran}
\usepackage{amsmath,amsfonts}
\usepackage{algorithmic}
\usepackage{algorithm}
\usepackage{array}
\usepackage[caption=false,font=normalsize,labelfont=sf,textfont=sf]{subfig}
\usepackage{textcomp}
\usepackage{stfloats}
\usepackage{url}
\usepackage{verbatim}
\usepackage{graphicx}
\usepackage{cite}
\usepackage{fontawesome5}
\usepackage{tcolorbox}
\usepackage{multirow}
\begin{document}

\title{ AI Ethics: An Empirical Study on the Views of Practitioners and Lawmakers}

\author{
\IEEEauthorblockN{Arif Ali Khan\IEEEauthorrefmark{1},
Muhammad Azeem Akbar\IEEEauthorrefmark{2},
Mahdi Fahmideh\IEEEauthorrefmark{3},
Peng Liang\IEEEauthorrefmark{4},
Muhammad Waseem\IEEEauthorrefmark{7},
Aakash Ahmad\IEEEauthorrefmark{5},
Mahmood Niazi\IEEEauthorrefmark{6} and
Pekka Abrahamsson\IEEEauthorrefmark{7} \\
\IEEEauthorrefmark{1}University of Oulu, Finland, 
\IEEEauthorrefmark{2}Lappeenranta-Lahti University of Technology, Finland, 
\IEEEauthorrefmark{3}University of Southern Queensland, Australia, 
\IEEEauthorrefmark{4}Wuhan University, China, 
\IEEEauthorrefmark{5}Lancaster University Leipzig, Germany, 
\IEEEauthorrefmark{6}King Fahd University of Petroleum and Minerals, Saudi Arabia, 
\IEEEauthorrefmark{7}University of Jyvaskyla, Finland
}
\thanks{Currently Under Review in IEEE Transactions XXXX}}



\maketitle

\begin{abstract}
Artificial Intelligence (AI) solutions and technologies are being increasingly adopted in smart systems context, however, such technologies are continuously concerned with ethical uncertainties. Various guidelines, principles, and regulatory frameworks are designed to ensure that AI technologies bring ethical well-being. However, the implications of AI ethics principles and guidelines are still being debated. To further explore the significance of AI ethics principles and relevant challenges, we conducted a survey of 99 representative AI practitioners and lawmakers (e.g., AI engineers, lawyers) from twenty countries across five continents. To the best of our knowledge, this is the first empirical study that encapsulates the perceptions of two different types of population (AI practitioners and lawmakers) and the study findings confirm that \textit{transparency}, \textit{accountability}, and \textit{privacy} are the most critical AI ethics principles. On the other hand, \textit{lack of ethical knowledge}, \textit{no legal frameworks}, and \textit{lacking monitoring bodies} are found the most common AI ethics challenges. The impact analysis of the challenges across AI ethics principles reveals that \textit{conflict in practice} is a highly severe challenge. Moreover, the perceptions of practitioners and lawmakers are statistically correlated with significant differences for particular principles (e.g. \textit{fairness}, \textit{freedom}) and challenges (e.g. \textit{lacking
monitoring bodies}, \textit{machine distortion}). Our findings stimulate further research, especially empowering existing capability maturity models to support the development and quality assessment of ethics-aware AI systems.
\end{abstract}


\section{Introduction}
\label{Introduction}
Artificial Intelligence (AI) becomes necessary across a vast array of industries including health, manufacturing, agriculture, and banking \cite{AR1}. AI technologies have the potential to substantially transform society and offer various technical and societal benefits, which are expected to happen from high-level productivity and efficiency. In line with this, the ethical guidelines presented by the
independent high-level expert group on artificial intelligence (AI HLEG) highlights that \cite{AR2}:

\faLightbulb{ “\textit{AI is not an end in itself, but rather a promising means to increase human flourishing, thereby enhancing individual and societal well-being and the common good, as well as bringing progress and innovation.}”}

However, the promising advantages of AI technologies have been considered with worries that the complex and opaque systems might bring more social harms than benefits \cite{AR1}. People start thinking beyond the operational capabilities of AI technologies and investigating the ethical aspects of developing strong and potentially life consequential technologies. For example, US government and many private companies do not use the virtual implications of decision-making systems in health, criminal justice, employment, and creditworthiness without ensuring that these systems are not coded intentionally or unintentionally with structural biases \cite{AR1}.

Concomitant with advances in AI systems, we witness the ethical failure scenarios. For example, a high rate of unsuccessful job applications that were processed by the Amazon recruitment system was later found biased in analysis of the selection criteria against women applicants and triggered discriminatory issues \cite{AR3}. Since decisions and recommendations made by AI systems may undergone people lives, the need for developing pertinent policies and principles addressing the ethical aspects of AI systems is crucial. Otherwise, the harms caused by AI systems will jeopardize the control, safety, livelihood and rights of people. AI systems are not only concerned with technical efforts, but also need to consider the social, political, legal, and intellectual aspects. However, AI's current state of ethics is broadly unknown to the public, practitioners, policy, and lawmakers \cite{AR4}\cite{vallor2017artificial}.

Extensively, the ethically aligned AI system should meet the following three components through the entire life cycle \cite{AR2}: 1) compliance with all the applicable laws and regulations, 2) adherence to ethical principles and values, and 3) technical and social robustness.
To the best of our knowledge, there is a dearth of empirical study to uncover the above core components in the view of industrial practitioners and lawmakers. For instance, as will be elaborated in Section 7, Vakkuri et al. \cite{AR4} conducted a survey study to determine industrial perceptions based only on four AI ethics principles. Lu et al.\cite{AR12} conducted interviews with researchers and practitioners to understand the AI ethics principles implications and the motivation for rooting these principles in the design practices. Similarly, Leikas et al. \cite{AR5} mainly focused on AI ethics guidelines. Given the lack of empirical studies exploring principles and challenges associated with AI ethics, we strive to answer the following research questions:

\begin{tcolorbox} [sharp corners, boxrule=0.1mm,]
\small
\textbf{RQ1}:What are the practitioners' and lawmakers' insights on AI ethics principles and challenges?
\end{tcolorbox}

\textbf{\underline{Rationale}}: \textbf{RQ1} aims to digest the perceptions of practitioners and lawmakers to empirically evaluate the systematic literature review (SLR) study based identified AI ethics principles and challenges \cite{AR13}. The answer to \textbf{RQ1} provides a better understanding of the most common AI ethics principles and challenges with respect to practitioners and lawmakers point of views.

\begin{tcolorbox} [sharp corners, boxrule=0.1mm,]
\small
\textbf{RQ2}: What would be the severity impacts of identified challenges across the AI ethics principles?
\end{tcolorbox}

\textbf{\underline{Rationale}}: \textbf{RQ2} aims to measure the severity impacts of challenging factors across AI ethics principles. The answer to \textbf{RQ2} would inform practitioners for the most severe challenges before initiating the AI ethics activities.

\begin{tcolorbox} [sharp corners, boxrule=0.1mm,]
\small
\textbf{RQ3}: How these challenges and principles are differently perceived by practitioners and lawmakers?

\end{tcolorbox}

\textbf{\underline{Rationale}}: The empirical data were collected from two types of populations (practitioners and lawmakers). The answer to \textbf{RQ3} would portray a better understanding of significant differences between the opinion of targeted populations for AI ethics principles and challenges.

To address these RQs, we conducted a survey study by encapsulating the views and opinions of practitioners and lawmakers regarding AI ethics principles and challenges by collecting data from 99 respondents across 20 different countries.

\section{Background}
\label{Background}
Generally, the AI ethics is classified under the umbrella of applied ethics, which mainly concerns with ethical issues associated with developing and using the AI systems. It focuses on linking how an AI system could raise worries related to human autonomy, freedom in a democratic society, and quality of life. Ethical reflection across AI technologies could serve in achieving multiple societal purposes \cite{AR2}. For instance, it can stimulate focusing on innovations that aim to foster ethical values and bring collective well-being. Ethically aligned or trustworthy AI technologies can flourish sustainable well-being in society by bringing prosperity, wealth maximization, and value creation \cite{AR2}.    

It is vital to understand the development, deployment, and use of AI technologies to ensure that everyone can build a better future and live a thriving life in the AI-based world. However, the increasing popularity of AI systems has raised concerns such as reliability and impartiality of decision-making scenarios \cite{AR2}. We need to make sure that decision-making support of AI technologies must have an accountable process to ensure that their actions are ethically aligned with human values that should not be compromised \cite{AR2}.

In this regard, different organizations and technology giants developed committees to draft the AI ethics guidelines. Google and SAP presented the guidelines and policies to develop ethically aligned AI systems \cite{AR6}. Similarly, the Association of Computing Machinery (ACM), Access Now, and Amnesty International jointly proposed the principles and guidelines to develop an ethically mature AI system \cite{AR6}. In Europe, the (AI HLEG) guidelines are developed for promoting trustworthy AI \cite{AR2}. The Ethically Aligned Design (EAD) guidelines are presented by IEEE, consisting of a set of principles and recommendations that focus on the technical and ethical values of AI systems \cite{AR7}. In addition, the joint ISO/IEC international standard committee proposed the ISO/IEC JTC 1/SC 42 standard which covers the entire AI ecosystem, including trustworthiness, computational approach, governance, standardization, and social concerns \cite{AR8}.

However, various researchers claim that the extant AI ethics guidelines and principles are not effectively adopted in industrial settings. McNamara et al. \cite{AR9} conducted an empirical study to understand the influence of the ACM code of ethics in the software engineering decision-making process. Surprisingly, the study findings reveal that no evidence has been found that the ACM code of ethics regulate decision-making activities. Vakkuri et al. \cite{AR10} conducted multiple interviews to know the status of ethical practices in the domain of the AI industry. The study findings uncover the fact that various guidelines are available; however, their deployment in industrial domains are far from being mature. The gap between AI ethics research and practice remains an ongoing challenge. To bridge this gap, we previously conducted an SLR study to provide a comprehensive and state-of-the-art overview of AI ethics principles and challenges \cite{AR13}. This study is extended based on the SLR findings \cite{AR13} to provide empirical insights to know the significance of AI ethics principles, challenges, and their impact by encapsulating the views of AI practitioners and lawmakers.
\section{Methodology} \label{SettingtheStage}
We deemed two groups of research participants would be relevant to this survey- AI practitioners and lawmakers.  On one hand, practitioners often make the design decisions and have higher ethical responsibilities compared to others. Practitioners often make the design decisions of complex autonomous systems with less ethical knowledge. The magnitude of risks in AI systems makes practitioners responsible for understanding ethical attributes. To achieve reliable outcomes, it is essential to know the practitioners understanding of AI ethics principles and challenges. 

On the other hand, law resolves everyday conflicts and sustains order in social life. People consider law an information source as it impacts social norms and values \cite{AR11}. The aim of considering this type of population (lawmakers) is to understand the application of the law to AI ethics. The data collected from legislation personnel will uncover the question, of whether standing AI ethics principles are sufficient, or is there a need for innovative standards \cite{AR11}? 

\begin{figure}[t]
 \centering
  \includegraphics[width=0.48\textwidth]{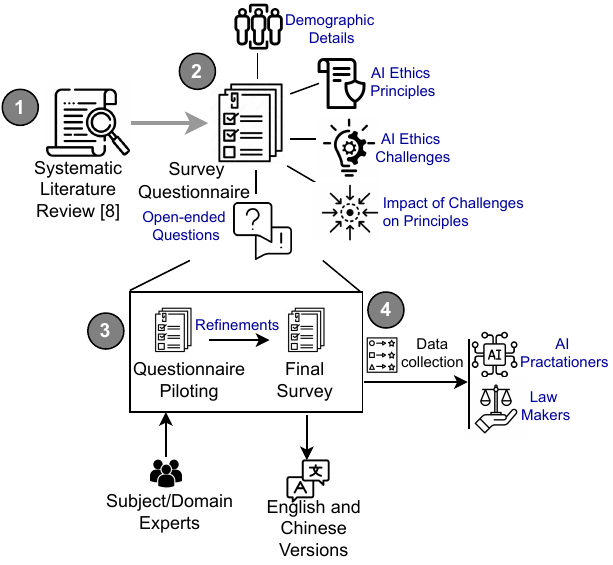}
\caption{Overview of the research methodology}
\label{fig:Research Methodology}
\end{figure}

We used industrial collaboration contacts to search the AI practitioners and sent a formal invitation to participate in this survey. Moreover, various law forums across the world were contacted and requested to participate in this study. The targeted populations were approached using social media networks including LinkedIn, WeChat, ResearchGate, Facebook, and personal email addresses. The overview of research methodology is depicted in Figure \ref{fig:Research Methodology}

The survey instrument consisted of four core sections: 1) demographics 2) AI ethics principles 3) challenges 4) challenges impact on principles. The survey questionnaire also includes open-ended questions to know the novel principles and challenges that were not identified in the SLR study \cite{AR13}. The Likert scale is used to evaluate the significance of each principle and challenge and assess the severity level of the challenging factors. The survey instrument is structured both in English and Chinese language. The software industry in China is flourishing like never before, where AI is taking the front seat and is home to some of the leading technology giants in the world, such as Huawei, Alibaba, Baidu, Tencent, and Xiaomi. However, it would be challenging to collect the data from the Chinese industry because of the language barriers. Mandarin is the national and official language in China, unlike India, where English is commonly used for official purposes. Therefore, the Chinese version of the survey instrument is developed to cover the major portion of the targeted population. Both English and Chinese versions of the survey instrument are available online for replication \cite{replication}. 

The piloting of the questionnaire is performed by inviting three external subject/domain experts. The experts' suggestions were mainly related to the overall design, and understandability of the survey questions. The suggested changes were incorporated, and the survey instrument was finalized based on the authors' consensus (see Figure \ref{fig:Research Methodology}). The final survey instrument was online deployed using Google forms (English version) and Tencent questionnaire (Chinese version). The first two authors engaged with the data collection process, while the next co-authors frequently monitored/screened the participants' responses. The data collection process was started in September 2021 and ended up in April 2022 with initial 107 total responses. It should be noted, we provided the consensus details in the information sheet of the survey questionnaire \cite{replication} and only considered the agreed responses for further analysis. 

The manual review revealed that eight responses were incomplete and we only considered 99 responses for the final data analysis.The third author mainly extracted and analysed the survey data. The descriptive data were analyzed using the frequency analysis approach. The frequency-based tables and charts are created for the identified AI ethics principles and challenges (see Section \ref{sec:Results}). Frequency analysis is more suitable for analyzing a group of variables and for both numeric and ordinal types of data \cite{bland2015introduction}. The significance of identified AI ethics principles and challenges is evaluated based on the level of agreement between the two types of populations (AI practitioners, lawmakers) (see Section \ref{sec:Statistical inferences (RQ3)}). The same data analysis approach has been used in different other similar nature of studies \cite{akbar2022srcmimm}\cite{khan2017systematic}\cite{niazi2016toward}. Finally, various Zoom consent meetings were called and invited all the authors to overview the study results and provide feedback. The study replication package is provided in \cite{replication}.

\section{Results and Discussions}\label{sec:Results}
We now present the final results and discussions of the survey findings based on the final agreement of all the authors, particularly (i)  demography details of survey participants, (ii)  survey participants’ perceptions of AI ethics principles and challenges, (iii)  severity impact of identified challenges across the AI ethics principles, and (IV) statistically significant differences between opinion of both type of populations (practitioners and lawmakers) for the identified principles and challenges.

\subsection{Demographic details}\label{sec:demographic}
\begin{figure*}
\centerline{\includegraphics[width=0.8\textwidth]{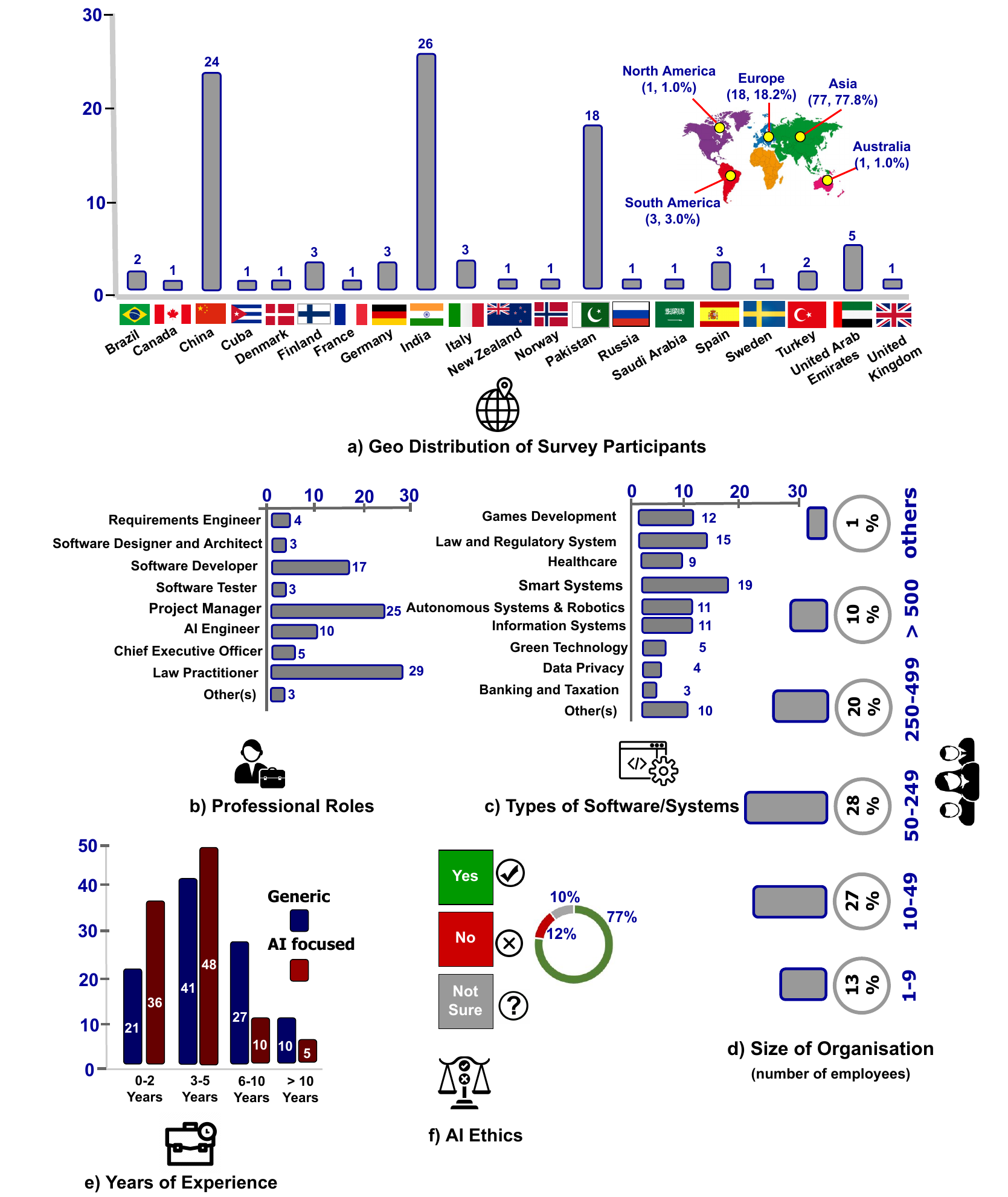}}
\caption{Demographic details of survey participants}
	\label{Fig:Demographics}
\end{figure*}

Frequency analysis was performed to organize the descriptive data and it is more suitable for analyzing a group of variables both for numeric and ordinal data. We noticed that 99 respondents from 20 countries across 5 continents with 9 roles and 10 different backgrounds participated in the survey study (see Figure \ref{Fig:Demographics}(a-c)). The organizational size (number of employees) of survey participants mostly ranges from \textit{50 to 249}, which is 28\% of the total responses (see Figure \ref{Fig:Demographics}(d)). Of all the responses, majority (48\%) have \textit{3-5 years} of experience working with AI focused projects as practitioners or lawmakers (see Figure \ref{Fig:Demographics}(e)). 

Participants were asked to explain their opinions about the perceived importance of AI systems in their organization. The majority of the participants positively agreed. For instance, 77\% mentioned that their respective organizations consider ethical aspects in AI processes or develop policies for AI projects, 12\% answered negatively, and 10\% were not sure about it (see Figure \ref{Fig:Demographics}(f)). We mapped the respondents’ roles across nine different categories using thematic mapping (see Figure \ref{Fig:Demographics}(b)). The final results show that the most of the respondents (29\%) are classified across the \textit{law practitioner} category. Similarly, the working domains of the participants’ organizations are conceptually framed in 10 core categories and the results revealed that most (19\%) of the organizations are working on \textit{smart systems} (see Figure \ref{Fig:Demographics}(c)).

\subsection{AI ethics principles and challenges (RQ1)} \label{sec:AI ethics principles and challenges (RQ1)}

\begin{figure*}
\centering
\includegraphics[width=0.95\textwidth]{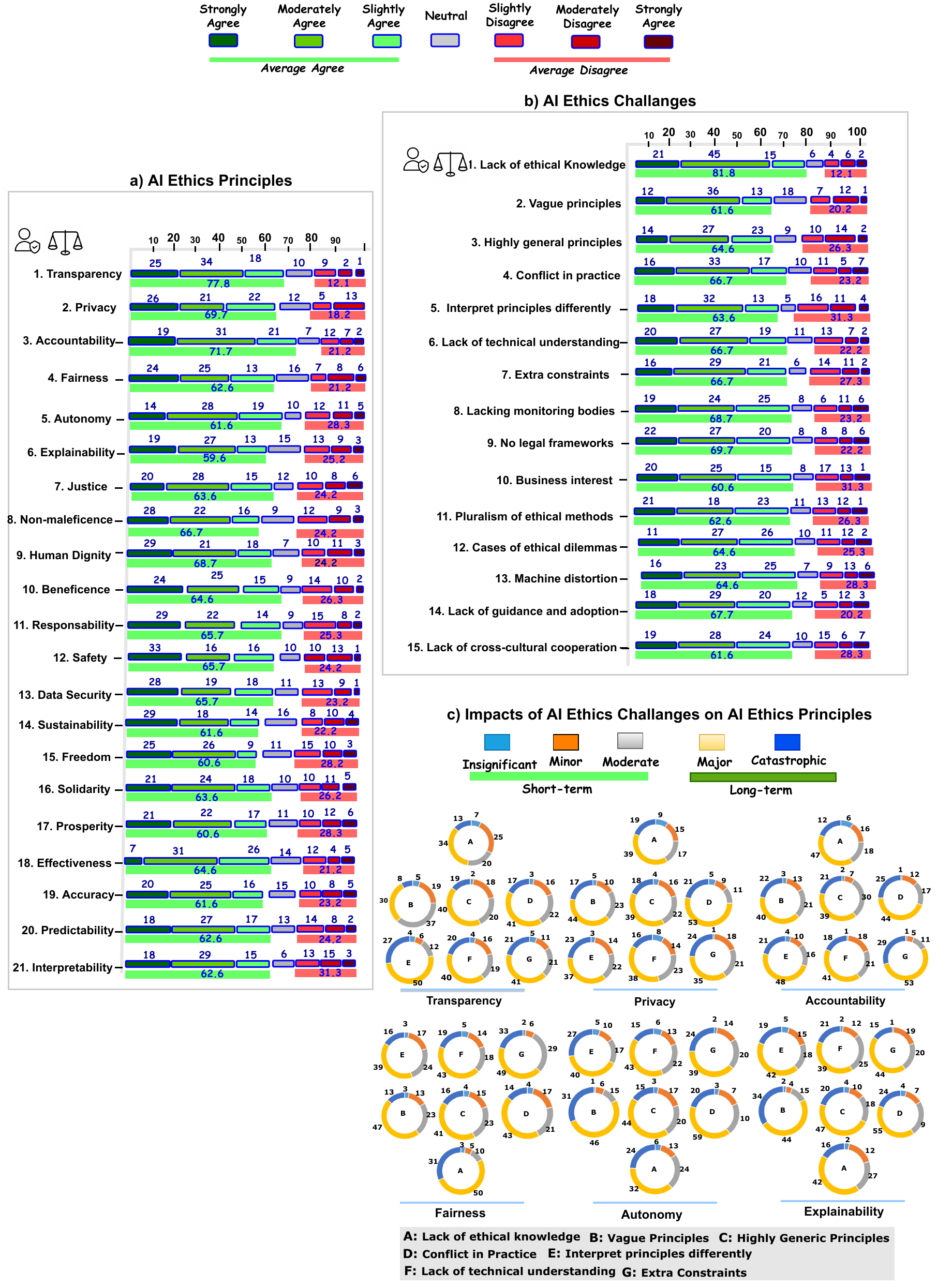}
 	\caption{Survey participants perceptions on AI ethics principles and challenges}
	\label{Fig:SurveyFindings}
\end{figure*}

The survey responses are classified as average agree, neutral and average disagree (see Figure \ref{Fig:SurveyFindings}(a-b)). We observed that (approx. 65\%) of the respondents positively confirmed the AI ethics principles and challenges identified in the SLR study \cite{AR13}. 

\subsubsection{AI ethics principles}
The results illustrate that the majority of the survey participants positively agreed (approx. 64\%) to consider the identified list of AI ethics principles (see Figure \ref{Fig:SurveyFindings}(a)). For instance, one survey participant mentioned that: 

\faComment{} “\textit{The listed AI ethics principles are comprehensive and extensive to cover various aspects of ethics in AI.}” 

We noticed that 77.8\% of survey respondents thought \textit{transparency} as the most significant AI ethics principle. This is an interesting observation as \textit{transparency} is equally confirmed as one of the core seven essential requirements by AI HLEG \cite{AR2} for realizing the ‘trustworthy AI’. \textit{Transparency} provides detailed explanations of logical AI models and decision-making structures understandable to the system stakeholders. Moreover, it deals with the public perceptions and understanding of how AI systems work. Broadly, it is a societal and normative ideal of “openness”.

The second most significant principle to the survey participants was \textit{accountability} (71.7\%). It refers to the expectations or requirements that the organizations or individuals need to ensure throughout the lifecycle of an AI system. They should be accountable according to their roles and applicable regulatory frameworks for the system design, development, deployment, and operation by providing documentation on the decision-making process or conducting regular auditing with proper justification. \textit{Privacy} is the third most frequently occurred principle, supported by 69.7\% of the survey participants. It refers to preventing harm, a fundamental right specifically affected by the decision-making system. \textit{Privacy} compels data governance throughout the system lifecycle, covering data quality, integrity, application domain, access protocols, and capability to process the data in a way that safeguards privacy. It must be ensured that the data collected and manipulated by the AI system shall not be used unlawfully or unfairly discriminate against human beings. For example, one of the respondents mentioned that 

\faComment{} “\textit{The privacy of hosted data used in AI applications and the risk of data breaches must be considered.}” 

In general, the survey findings of AI ethics principles are confirmatory to the widely adopted accountability, responsibility, and transparency (ART) framework \cite{AR15} and the findings of an industrial empirical study conducted by Ville et al. \cite{AR10}. Both studies \cite{AR10} \cite{AR15} jointly considered \textit{transparency} and \textit{accountability} as the core AI ethics principles, which is consistent with the findings in this survey. On contrary, we noticed that \textit{privacy} has been ignored in both mentioned studies \cite{AR10} \cite{AR15}, but is placed as the third most significant principle in this survey. The reason might be that, as more and more AI systems have been placed online, the significance of \textit{privacy} and data protection is increasingly recognized \cite{bansal2022internet}. Presently, various countries embarked on legislation to ensure the protection of data and \textit{privacy}.

\subsubsection{AI ethics challenges}

Further, the results reveal that the majority of the survey respondents (approx. 66\%) confirmed the identified challenging factors \cite{AR13} (see Figure \ref{Fig:SurveyFindings}(b)). \textit{Lack of ethical knowledge} is considered as the most frequently cited challenge by (81.8\%) of the survey participants. It exhibits that knowledge of ethical aspects across AI systems is largely ignored in industrial settings. There is a significant gap between research and practice in AI ethics. Extant guidelines and policies devised by researchers and regulatory bodies discussed different ethical goals for AI systems. However, these goals have not been widely adopted in the industrial domain because of limited knowledge of scaling them in practice. The findings are in agreement with the results of industrial study conducted by Ville et al. \cite{AR10}, concluding that ethical aspects of AI systems are not exclusively considered, and it mainly happened because of a lack of knowledge, awareness, and personal commitment. We noticed that \textit{no legal frameworks} (69.7\%) is ranked as the second most common challenge for considering ethics in the AI domain. The proliferation of AI technologies in high-risk areas starts mounting the pressure of designing ethical and legal standards and frameworks to govern them \cite{cath2018governing}. It highlights the nuances of the debate on AI law and lays the groundwork for a more inclusive AI governance framework \cite{schiff2020s}. The framework shall focus on most pertinent ethical issues raised by the AI systems, the use of AI across industry and government organisations, and economic displacement (i.e. the ethical reply to the loss of jobs as a result of AI-based automation). The third most common challenging factor is \textit{lacking monitoring bodies}, and it was highlighted by (68.7\%) of the survey participants. \textit{Lacking monitoring bodies} refers to the lack of regulatory oversight to assess ethics in AI systems \cite{AR13}. It raises the issue of public bodies’ empowerment to monitor, and audit the enforcement of ethical concerns in AI technologies by the domain (e.g., health, transport, education). One survey respondent mentioned that 

\faComment{} “\textit{I believe it shall be mandatory for the industry to get standard approval from monitoring bodies to consider ethics in the development process of AI systems.}” 

Monitoring bodies extensively promote and observe the ethical values in society and evaluate technology development associated with ethical aspects of AI \cite{AR2}. They would be tasked to advocate and define responsibilities and develop rules, regulations, and practices in a situation where the system takes a decision autonomously. The monitoring group should ensure “ethics by, in and for design” as mentioned in AI HLEG \cite{AR2} guidelines. 

Additionally, the survey participants elaborated on new challenging factors. For instance, one of the participants mentioned that 

\faComment{}  “\textit{Implicit biases in AI algorithms such as data discrimination and cognitive biases could impact system transparency.}” 

Similarly, the other respondent reported that 

\faComment{} “\textit{Biases in the AI system’s design might bring distress to a group of people or individuals.}” 

Moreover, a survey respondent explicitly considered the \textit{lack of tools for ethical transparency} and \textit{AI biases} as significant challenges to AI ethics. We noticed that \textit{AI biases} is reported as the most common additional challenge. It will be interesting to further explore (i) the type of biases that might be embedded with the AI algorithms, (ii) the causes of these biases, and (iii) corresponding countermeasures to minimize the negative impact on AI ethics.

\subsection{Severity impacts of identified challenges (RQ2)} \label{sec:Severity impacts of identified challenges (RQ2)}

We selected the most frequently reported seven challenging factors and six principles discussed in our SLR study \cite{AR13}. The aim is to investigate the severity impact of the seven challenges \textit{(i.e., lack of ethical knowledge, vague principles, highly general principles, conflict in practice, interpret principles differently, lack of technical understanding, and extra constraints)} across the six AI ethics principles \textit{(i.e. transparency, privacy, accountability, fairness, autonomy, and explainability)}. The survey participants were asked to rate the severity impact using the Likert scale: short-term (insignificant, minor, moderate) and long-term (major, and catastrophic)  (see Figure \ref{Fig:SurveyFindings}(c)). The results revealed that most challenges have long-term impacts on the principles (major, and catastrophic). 

For the \textit{transparency} principle, we noticed that the challenging factor \textit{interpret principles differently} has significant long-term impacts, and 77\% (i.e., 50\% major, and 27\% catastrophic) of the survey participants agreed to it. The interpretation of ethical concepts can change for a group of people and individuals. For instance, the practitioners might perceive \textit{transparency} differently (more focused on technical aspects) than law and policymakers, who have broad social concerns. Furthermore, \textit{lack of ethical knowledge} has a short-term impact on the \textit{transparency} principle, and it is evident from the survey findings supported by 52\% (7\% insignificant, 25\% minor, and 20\% moderate) responses. Lack of knowledge could be instantly covered by attaining knowledge, understanding, and awareness of transparency concepts. 

\textit{Conflict in practice} is deemed the most significant challenge to the \textit{privacy} principle. Hence, 74\% (i.e., 53\% major, and 21\% catastrophic) survey respondents considered it a long-term severe challenge. Various groups, organizations, and individuals might have opinion conflicts associated with \textit{privacy} in AI ethics \cite{stahl2018ethics}. It is critical to interpret and understand privacy conflicts in practice. We noticed that (82\%) of survey participants considered the challenging factor \textit{extra constraints} as the most severe (long-term) challenge for both \textit{accountability} and \textit{fairness} principles. Situational constraints, including organizational politics, lack of information, and management interruption, could possibly interfere with the accountability and fairness measures \cite{krijger2021enter}. It could negatively impact the employee’s motivation and interest to explicitly consider ethical aspects across the AI activities. Interestingly, (79\%) of the survey respondents considered \textit{conflict in practice} as the most common (long-term) challenge for \textit{autonomy} and \textit{explainability} principles.

Overall, we could interpret that \textit{conflict in practice}  is the most severe challenge, and its average occurrence is $>$60\% for all the principles. It gives a general understanding to propose specific solutions that focus on tackling the opinion conflict regarding the real-world implication of AI ethics principles. The results further reveal \textit{lack of ethical knowledge} has an average (28\%) short-term impact across selected AI ethics principles. The lack of knowledge gap could be covered by conducting training sessions, workshops, certification, and encouraging social awareness of AI ethics \cite{AR13}. Knowledge increases the possibility of AI ethics success and acceptance in the best practice of the domain. 

\subsection{Statistical inferences (RQ3)} \label{sec:Statistical inferences (RQ3)}

We performed non-parametric statistical analysis \cite{myers2004spearman}\cite{de2016comparing} to evaluate the significant differences  and similarities between the opinion of lawmakers and software practitioners. The same non-parametric statistical analysis are previously performed in different other similar nature of studies  \cite{akbar2022srcmimm}\cite{khan2020systematic}\cite{khan2017systematic}. The frequency-based ranking of both datasets is identified for AI ethics principles (see Table \ref{tab:Rank-Principles}) and challenges (see Table  \ref{tab:AI ethics challenges ranks}) to set common measures for non-parametric Spearman's Rank-order correlation coefficient. It gives the linear  dependence between a set of variables, ranging from (rs (co-relation coefficient) = +1 to -1), where +1 indicates a total linear dependency \cite{myers2004spearman}\cite{de2016comparing}.

\subsubsection{Significant differences for AI ethics principles}

\begin{figure}[]
\centering
\includegraphics[width=0.4\textwidth]{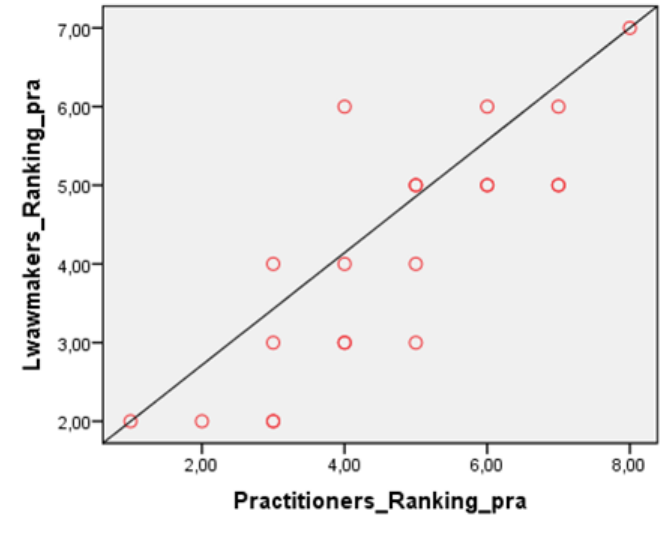} 
 	\caption{Scatter plot of ranks for AI ethics principles}
	\label{Fig:Scatter-Ranks-Principles}
\end{figure}

The Spearman's Rank-order correlation test was applied to statistically evaluate the significant differences between the practitioners and lawmakers perceptions on AI ethics principles. We obtained the Spearman's Rank-order correlation coefficient value (rs$=$0.819), which is statistically significant (p$=$0.000) (see Table \ref{tab:correlation_Principles}). The value (rs$=$0.819) and the scatter plot given in Figure \ref{Fig:Scatter-Ranks-Principles} show the strong correlation between the ranks of both datasets (lawmakers and software practitioners). The identified principles are widely discussed across multiple AI ethics guidelines, and it might be the reason why both practitioners and lawmakers equally agreed with the significance and implications of these principles. For example, \textit{transparency} is a common AI ethics principle, and practitioners and lawmakers ranked it in the first position. However, we also noticed significant differences (p$=$0.000) between both types of the population. For instance, lawmakers ranked \textit{fairness} at position five as the most important principle; however, the software practitioners placed it at position seven. It shows that fairness across AI activities is relatively important based on lawmakers' perceptions. It is because \textit{fairness} is a non-technical and more socially used term. Laws like EU GDPR impose concrete requirements on AI development organizations to safeguard fairness in AI system design, deployment, and data processing \cite{GDPR}. The low-ranked placement of \textit{fairness} by the practitioners might be because of limited knowledge and understanding of interpreting fairness technically, e.g., fairness in AI by design.

\begin{table*}[htbp] 
\centering
\caption{AI ethics principles ranks}
\label{tab:Rank-Principles}
\begin{tabular}{|l|c|c|c|c|}
\hline
\multicolumn{1}{|c|}{\textbf{Principles}} & \multicolumn{1}{c|}{\textbf{\begin{tabular}[c]{@{}c@{}}Lawmaker\\ (n=29)\end{tabular}}} & \multicolumn{1}{c|}{\textbf{\begin{tabular}[c]{@{}c@{}}Rank by\\ Lawmakers\end{tabular}}} & \multicolumn{1}{c|}{\textbf{\begin{tabular}[c]{@{}c@{}}Practitioners\\ (n=52)\end{tabular}}} & \multicolumn{1}{c|}{\textbf{\begin{tabular}[c]{@{}c@{}}Rank by\\ Practitioners\end{tabular}}} \\ \hline
Transparency & 25 & 1 & 52 & 1 \\ \hline
Privacy & 23 & 3 & 43 & 4 \\ \hline
Accountability & 24 & 2 & 44 & 3 \\ \hline
Fairness & 21 & 5 & 39 & 7 \\ \hline
Autonomy & 20 & 6 & 39 & 7 \\ \hline
Explainability & 19 & 7 & 38 & 8 \\ \hline
Justice & 20 & 6 & 41 & 5 \\ \hline
Non-maleficence & 23 & 3 & 41 & 5 \\ \hline
Human dignity & 24 & 2 & 44 & 3 \\ \hline
Beneficence & 22 & 4 & 44 & 3 \\ \hline
Responsibility & 22 & 4 & 43 & 4 \\ \hline
Safety & 23 & 3 & 44 & 3 \\ \hline
Data Security & 22 & 4 & 41 & 5 \\ \hline
Sustainability & 24 & 2 & 45 & 2 \\ \hline
Freedom & 21 & 5 & 39 & 7 \\ \hline
Solidarity & 21 & 5 & 41 & 5 \\ \hline
Prosperity & 20 & 6 & 40 & 6 \\ \hline
Effectiveness & 20 & 6 & 43 & 4 \\ \hline
Accuracy & 21 & 5 & 40 & 6 \\ \hline
Predictability & 23 & 3 & 43 & 4 \\ \hline
Interpretability & 21 & 5 & 40 & 6 \\ \hline
\end{tabular}%
\end{table*}

\begin{table*}[htbp]
\centering
\caption{Practitioners and lawmakers perceptions co-relation for AI ethics principles}
\label{tab:correlation_Principles}
\begin{tabular}{|lllrr|}
\hline
\multicolumn{3}{|l|}{} & \multicolumn{1}{c|}{\begin{tabular}[c]{@{}c@{}}Lwawmakers\_\\ Ranking\_pra\end{tabular}} & \multicolumn{1}{c|}{\begin{tabular}[c]{@{}c@{}}Practitioners\_\\ Ranking\_pra\end{tabular}} \\ \hline
\multicolumn{1}{|l|}{\multirow{6}{*}{Spearman's rho}} & \multicolumn{1}{c|}{\multirow{3}{*}{\begin{tabular}[c]{@{}c@{}}Lwawmakers\_\\ Ranking\_pra\end{tabular}}} & \multicolumn{1}{l|}{Correlation Coefficient} & \multicolumn{1}{r|}{1.000} & 0.819** \\ \cline{3-5} 
\multicolumn{1}{|l|}{} & \multicolumn{1}{c|}{} & \multicolumn{1}{l|}{Sig. (2-tailed)} & \multicolumn{1}{r|}{.} & 0.000 \\ \cline{3-5} 
\multicolumn{1}{|l|}{} & \multicolumn{1}{c|}{} & \multicolumn{1}{l|}{N} & \multicolumn{1}{r|}{21} & 21 \\ \cline{2-5} 
\multicolumn{1}{|l|}{} & \multicolumn{1}{l|}{\multirow{3}{*}{\begin{tabular}[c]{@{}l@{}}Practitioners\_\\ Ranking\_pra\end{tabular}}} & \multicolumn{1}{l|}{Correlation Coefficient} & \multicolumn{1}{r|}{0.819**} & 1.000 \\ \cline{3-5} 
\multicolumn{1}{|l|}{} & \multicolumn{1}{l|}{} & \multicolumn{1}{l|}{Sig. (2-tailed)} & \multicolumn{1}{r|}{0.000} & . \\ \cline{3-5} 
\multicolumn{1}{|l|}{} & \multicolumn{1}{l|}{} & \multicolumn{1}{l|}{N} & \multicolumn{1}{r|}{21} & 21 \\ \hline
\multicolumn{5}{|l|}{**. Correlation is significant at the 0.01 level (2-tailed).} \\ \hline
\end{tabular}%
\end{table*}

In addition to Spearman's Rank order co-relation, we also applied the independent t-test to compare the mean differences of the ranks obtained from both types of population (see Table \ref{tab:ttest_principples} and Table \ref{tab:groupstatistics_principles}). Since Levene's Test is slightly significant (i.e., p$=$0.051$>$0.05), therefore, we assume that the variances are approximately equal. Based on this assumption, the results of t-test (i.e., t $=$ 0.942, p $=$ 0.661 $>$ 0.05) show that there is no high-level significant differences between both variables. The results show that the degree of agreement between lawmakers and practitioners concerning AI ethics principles is positive. It means that both populations (lawmakers and software practitioners) equally consider the importance of AI ethics principles. The group statistics for both variables are given in Table \ref{tab:groupstatistics_principles}.

\begin{table*}[]
\centering
\caption{Independent samples t-test of AI ethics principles}
\label{tab:ttest_principples}
\begin{tabular}{|ll|cc|ccccccr|}
\hline
\multicolumn{2}{|l|}{\multirow{3}{*}{}} & \multicolumn{2}{c|}{\begin{tabular}[c]{@{}c@{}}Levene's Test \\ for Equality \\ of Variances\end{tabular}} & \multicolumn{7}{c|}{t-test for Equality of Means} \\ \cline{3-11} 
\multicolumn{2}{|l|}{} & \multicolumn{1}{c|}{\multirow{2}{*}{F}} & \multirow{2}{*}{Sig.} & \multicolumn{1}{c|}{\multirow{2}{*}{t}} & \multicolumn{1}{c|}{\multirow{2}{*}{df}} & \multicolumn{1}{c|}{\multirow{2}{*}{\begin{tabular}[c]{@{}c@{}}Sig. \\ (2-tailed)\end{tabular}}} & \multicolumn{1}{c|}{\multirow{2}{*}{\begin{tabular}[c]{@{}c@{}}Mean \\ Difference\end{tabular}}} & \multicolumn{1}{c|}{\multirow{2}{*}{\begin{tabular}[c]{@{}c@{}}Std. Error \\ Difference\end{tabular}}} & \multicolumn{2}{c|}{\begin{tabular}[c]{@{}c@{}}95\% Confidence \\ Interval of the \\ Difference\end{tabular}} \\ \cline{10-11} 
\multicolumn{2}{|l|}{} & \multicolumn{1}{c|}{} &  & \multicolumn{1}{c|}{} & \multicolumn{1}{c|}{} & \multicolumn{1}{c|}{} & \multicolumn{1}{c|}{} & \multicolumn{1}{c|}{} & \multicolumn{1}{c|}{Lower} & \multicolumn{1}{c|}{Upper} \\ \hline
\multicolumn{1}{|l|}{\multirow{2}{*}{Rank}} & \begin{tabular}[c]{@{}l@{}}Equal variances\\ assumed\end{tabular} & \multicolumn{1}{r|}{.117} & \multicolumn{1}{r|}{0.051} & \multicolumn{1}{r|}{0.942} & \multicolumn{1}{r|}{40} & \multicolumn{1}{r|}{.661} & \multicolumn{1}{r|}{-.23810} & \multicolumn{1}{r|}{.53875} & \multicolumn{1}{r|}{-1.32695} & .85076 \\ \cline{2-11} 
\multicolumn{1}{|l|}{} & \begin{tabular}[c]{@{}l@{}}Equal variances \\ not assumed\end{tabular} & \multicolumn{1}{l|}{} & \multicolumn{1}{l|}{} & \multicolumn{1}{r|}{0.942} & \multicolumn{1}{r|}{39.616} & \multicolumn{1}{r|}{.661} & \multicolumn{1}{r|}{-.23810} & \multicolumn{1}{r|}{.53875} & \multicolumn{1}{r|}{-1.32727} & .85108 \\ \hline
\end{tabular}%
\end{table*}

\begin{table*}[]
\centering
\caption{AI ethic principles group statistics}
\label{tab:groupstatistics_principles}
\begin{tabular}{|l|l|c|c|c|c|}
\hline
 & Group & N & Mean & Std. Deviation & Std. Error Mean \\ \hline
\multirow{2}{*}{Rank} & lawmakers & 21 & 4.3810 & 1.65759 & .36172 \\ \cline{2-6} 
 & practitioners & 21 & 4.6190 & 1.82965 & .39926 \\ \hline
\end{tabular}%
\end{table*}

\subsubsection{Significant differences for AI ethics challenges} \label{sec: Significant differences for AI ethics challenges}
\begin{figure}
\centering
\includegraphics[width=0.4\textwidth]{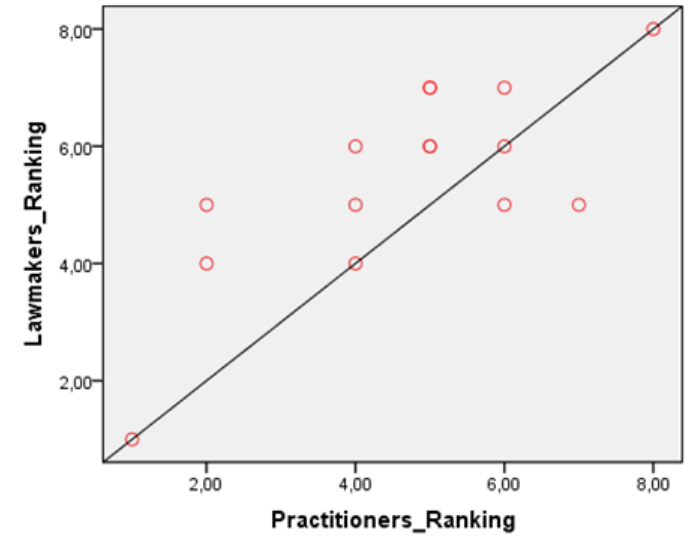} 
 	\caption{Scatter plot of ranks for AI ethics challenges}
	\label{Fig:Scatter-Ranks-challenges}
\end{figure}

Similar to AI ethics principles, the identified challenges are ranked (see Table \ref{tab:AI ethics challenges ranks}) and applied Spearman's rank-order correlation coefficient test to measure the significant differences. The correlation coefficient value (rs$=$0.628) shows a positive and statistically significant (p$=$0.012) correlation between both types of population (see Table \ref{tab:corr_challenges}). It indicate a moderate and statistically significant agreement between the opinions of lawmakers and practitioners concerning the AI ethics challenges (see Figure \ref{Fig:Scatter-Ranks-challenges}). For example, \textit{lacking monitoring bodies} is ranked second by the practitioners and fifth by the lawmakers. The practitioners mainly engage in team-oriented activities and are more concerned about human bias \cite{BiasAI}. Continuous socio-technical monitoring ensures delivering reliable, unbiased and fair outcomes. Avoiding proper monitoring deems to bring high ethical harm to practitioners and increase reputational risk \cite{BiasAI}.

\begin{table*}[]
\centering
\caption{AI ethics challenges ranks}
\label{tab:AI ethics challenges ranks}
\begin{tabular}{|l|l|l|l|l|}
\hline
\multicolumn{1}{|c|}{\textbf{Challenges}} & \multicolumn{1}{c|}{\textbf{\begin{tabular}[c]{@{}c@{}}Lawmaker\\ (n=29)\end{tabular}}} & \multicolumn{1}{c|}{\textbf{\begin{tabular}[c]{@{}c@{}}Rank by\\ Lawmakers\end{tabular}}} & \multicolumn{1}{c|}{\textbf{\begin{tabular}[c]{@{}c@{}}Practitioners\\ (n=52)\end{tabular}}} & \multicolumn{1}{c|}{\textbf{\begin{tabular}[c]{@{}c@{}}Rank by\\ Practitioners\end{tabular}}} \\ \hline
Lack of ethical Knowledge & 28 & 1 & 52 & 1 \\ \hline
Vague principles & 20 & 7 & 41 & 5 \\ \hline
Highly general principles & 22 & 6 & 41 & 5 \\ \hline
Conflict in practice & 22 & 6 & 43 & 4 \\ \hline
Interpret principles differently & 22 & 6 & 41 & 5 \\ \hline
Lack of technical understanding & 23 & 5 & 43 & 4 \\ \hline
Extra constraints & 23 & 5 & 40 & 6 \\ \hline
Lacking monitoring bodies & 23 & 5 & 45 & 2 \\ \hline
No legal frameworks & 24 & 4 & 45 & 2 \\ \hline
Business interest & 19 & 8 & 37 & 8 \\ \hline
Pluralism of ethical methods & 22 & 6 & 40 & 6 \\ \hline
Cases of ethical dilemmas & 20 & 7 & 40 & 6 \\ \hline
Machine distortion & 23 & 5 & 38 & 7 \\ \hline
Lack of guidance and adoption & 24 & 4 & 43 & 4 \\ \hline
Lack of cross-cultural cooperation & 20 & 7 & 41 & 5 \\ \hline
\end{tabular}%
\end{table*}

We also applied the independent t-test (see Table \ref{tab:ttest-challenges}, Table \ref{tab:group-stat-challenges}) to assess the mean differences between both types of the population with respect to AI ethics challenges. The calculated significance value of Levene's Test is (p$=$0.051$>$0.05); therefore, we assume the variances equally (see Table \ref{tab:ttest-challenges}). The t-test results, assuming equal variances (t$=$1.291, p$=$0.207$>$0.05), show that practitioners and lawmakers consider the signaficance of identified challenges equally. We could suppose that practitioners and lawmakers are equally aware of the reported challenges and understand their importance. The group statistics for both variables are provided in Table \ref{tab:group-stat-challenges}.
 
\begin{table*}[]
\centering
\caption{Practitioners and lawmakers perceptions co-relation for AI ethics challenges}
\label{tab:corr_challenges}
\begin{tabular}{|lllll|}
\hline
\multicolumn{5}{|l|}{Correlations} \\ \hline
\multicolumn{3}{|l|}{} & \multicolumn{1}{l|}{Lawmakers\_Ranking} & Practitioners\_Ranking \\ \hline
\multicolumn{1}{|l|}{\multirow{6}{*}{Spearman's rho}} & \multicolumn{1}{l|}{\multirow{3}{*}{Lawmakers\_Ranking}} & \multicolumn{1}{l|}{Correlation Coefficient} & \multicolumn{1}{l|}{1.000} & 0.628* \\ \cline{3-5} 
\multicolumn{1}{|l|}{} & \multicolumn{1}{l|}{} & \multicolumn{1}{l|}{Sig. (2-tailed)} & \multicolumn{1}{l|}{.} & 0.012 \\ \cline{3-5} 
\multicolumn{1}{|l|}{} & \multicolumn{1}{l|}{} & \multicolumn{1}{l|}{N} & \multicolumn{1}{l|}{15} & 15 \\ \cline{2-5} 
\multicolumn{1}{|l|}{} & \multicolumn{1}{l|}{\multirow{3}{*}{Practitioners\_Ranking}} & \multicolumn{1}{l|}{Correlation Coefficient} & \multicolumn{1}{l|}{0.628*} & 1.000 \\ \cline{3-5} 
\multicolumn{1}{|l|}{} & \multicolumn{1}{l|}{} & \multicolumn{1}{l|}{Sig. (2-tailed)} & \multicolumn{1}{l|}{0.012} & . \\ \cline{3-5} 
\multicolumn{1}{|l|}{} & \multicolumn{1}{l|}{} & \multicolumn{1}{l|}{N} & \multicolumn{1}{l|}{15} & 15 \\ \hline
\multicolumn{5}{|l|}{*.   Correlation is significant at the 0.05 level (2-tailed).} \\ \hline
\end{tabular}%
\end{table*}

\begin{table*}[]
\centering
\caption{Independent samples t-test of AI ethics challenges}
\label{tab:ttest-challenges}
\resizebox{\textwidth}{!}{%
\begin{tabular}{|ll|ll|lllllll|}
\hline
\multicolumn{2}{|l|}{\multirow{3}{*}{}} & \multicolumn{2}{l|}{\begin{tabular}[c]{@{}l@{}}Levene's Test   for  \\    \\ Equality of   Variances\end{tabular}} & \multicolumn{7}{l|}{t-test for   Equality of Means} \\ \cline{3-11} 
\multicolumn{2}{|l|}{} & \multicolumn{1}{l|}{\multirow{2}{*}{F}} & \multirow{2}{*}{Sig.} & \multicolumn{1}{l|}{\multirow{2}{*}{t}} & \multicolumn{1}{l|}{\multirow{2}{*}{df}} & \multicolumn{1}{l|}{\multirow{2}{*}{\begin{tabular}[c]{@{}l@{}}Sig. (2-   \\    \\ tailed)\end{tabular}}} & \multicolumn{1}{l|}{\multirow{2}{*}{\begin{tabular}[c]{@{}l@{}}Mean  \\    \\ Difference\end{tabular}}} & \multicolumn{1}{l|}{\multirow{2}{*}{Std. Error  Difference}} & \multicolumn{2}{l|}{\begin{tabular}[c]{@{}l@{}}95\% Confidence  \\    \\ Interval of the    \\    \\ Difference\end{tabular}} \\ \cline{10-11} 
\multicolumn{2}{|l|}{} & \multicolumn{1}{l|}{} &  & \multicolumn{1}{l|}{} & \multicolumn{1}{l|}{} & \multicolumn{1}{l|}{} & \multicolumn{1}{l|}{} & \multicolumn{1}{l|}{} & \multicolumn{1}{l|}{Lower} & Upper \\ \hline
\multicolumn{1}{|l|}{\multirow{2}{*}{Rank}} & Equal variances  assumed & \multicolumn{1}{l|}{.519} & .477 & \multicolumn{1}{l|}{1.291} & \multicolumn{1}{l|}{28} & \multicolumn{1}{l|}{.207} & \multicolumn{1}{l|}{.86667} & \multicolumn{1}{l|}{.67141} & \multicolumn{1}{l|}{-.50866} & 2.24199 \\ \cline{2-11} 
\multicolumn{1}{|l|}{} & Equal variances  not assumed & \multicolumn{1}{l|}{} &  & \multicolumn{1}{l|}{1.291} & \multicolumn{1}{l|}{27.145} & \multicolumn{1}{l|}{.208} & \multicolumn{1}{l|}{.86667} & \multicolumn{1}{l|}{.67141} & \multicolumn{1}{l|}{-.51061} & 2.24395 \\ \hline
\end{tabular}%
}
\end{table*}

\begin{table*}[]
\centering
\caption{AI ethic challenges group statistics}
\label{tab:group-stat-challenges}
\begin{tabular}{|l|l|l|l|l|l|}
\hline
 & group & N & Mean & Std. Deviation & Std. Error Mean \\ \hline
\multirow{2}{*}{Rank} & lawmakers & 15 & 5.1333 & 1.99523 & .51517 \\ \cline{2-6} 
 & practitioners & 15 & 4.2667 & 1.66762 & .43058 \\ \hline
\end{tabular}%
\end{table*}
Overall, we believe that practitioners and lawmakers are on the same page in considering AI ethics principles and challenges. 
However, for AI ethics challenges, the perceptions of practitioners and lawmakers are slightly different. 
We noticed that various AI ethics principles and guidelines are released in private and public sectors, which are very abstract, and incoherent for various stakeholders to implement \cite{munn2022uselessness}. The challenges of interpreting these vague principles are different with respect to the targeted group of stakeholders e.g., industrial and legislation perspectives \cite{munn2022uselessness}. In conclusion, there is a gap between high-minded principles and industrial practices, which needs alternative approaches based on mutual industrial and legislation consensus.

\section{Key findings and implications}
We now outline the key findings of the study to answer the RQs - AI ethics principles, challenges, severity impact and the statistically significant differences between the perceptions of practitioners and lawmakers. We also report the research and practical implications of the study findings.

\subsection{Summary and interpretation of the key findings}
The results of each RQ are thoroughly discussed in Section \ref{sec:Results} and the summary of core findings is presented in Table \ref{tab:Summary of key findings}, addressing RQ1 - confirming the identified AI ethics principles and challenges, RQ2 - measuring the severity impacts of the challenges across principles and RQ3 - practitioners and lawmakers perceptions of AI ethics principles and challenges. For RQ1, the summary of the study results highlights that both practitioners and lawmakers empirically confirm the AI ethics principles and challenges identified in our recent SLR study \cite{AR13}. We noticed that ($\geq$ 60\%) of the survey participants agreed to consider the reported principles and challenges. They further define the ranks of identified principles and challenges across a five-point Likert scale, which indicates that \textit{transparency}, \textit{accountability} and \textit{privacy} are the most critical principles \cite{AR15}\cite{AR10}; on the other hand, \textit{lack of ethical knowledge}, \textit{no legal frameworks} and \textit{monitoring bodies} appeared as the most frequently occurred challenging factors. Similarly, the summary of findings to address RQ2 reveal that \textit{conflict in practice} emerged as the most severe challenging factor for the identified AI ethics principles (see Table \ref{tab:Summary of key findings}). For certain cases, the AI ethics principles come into conflict, and their practical values become unrealistic- prioritizing one might inadvertently compromise another \cite{whittlestone2019role}. Whittlestone et al. \cite{whittlestone2019role} argue that thorough exploration is required to encounter and articulate the conflict and tensions across the AI ethics principles. For RQ3, we noticed that the perceptions of both types of populations (practitioners, lawmakers) are correlated and statistically significant for specific challenges. It is because the existing principles are too vague, generic, conceptual and no match for the specific and complex AI problems. Stakeholders have different perceptions of the challenges raised because of implementing the generic AI ethics principles. A broader consensus of multiple stakeholders- practitioners, lawmakers, and regulatory bodies required to define domain-specific principles and guidelines \cite{whittlestone2019role}. Overall, the summary of the findings in Table \ref{tab:Summary of key findings} is self-explanatory and encapsulates the core results discussed in Section \ref{sec:Results}.

\begin{table*}
\centering
\caption{Summary of the key findings}
\label{tab:Summary of key findings}
\begin{tabular}{|m{3cm}|m{10cm}|}
\hline
\textbf{Research Question} & \textbf{Summary} \\ \hline
RQ1: What are the practitioners' and lawmakers' insights on AI ethics principles and challenges? & 
\begin{itemize}
    \item Practitioners and lawmakers empirically confirm the principles and challenges reported in our previously published SLR study \cite{AR13}.
    \item \textit{Transparency}, \textit{accountability}, and \textit{privacy} emerged as the most perceived principles.
    \item \textit{Lack of ethical knowledge}, \textit{no legal frameworks}, and \textit{lacking monitoring bodies} appeared to be the most frequently cited challenging factors.
\end{itemize} \\
\hline
RQ2: What would be the severity impacts of identified challenges across the AI ethics principles? & 
Transparency \vspace{0mm}
\begin{itemize}
    \item \textit{Interpret principles differently} has long-term impacts (50\% major, 27\% catastrophic)
    \item \textit{Lack of ethical knowledge} has short-term impacts (7\% insignificant, 25\% minor, 20\% moderate)
\end{itemize} 
Privacy \vspace{0mm}
\begin{itemize}
    \item \textit{Conflict in practice} has long-term impacts (53\% major, 21\% catastrophic)
    \item \textit{Lack of technical understanding} has short-term impacts (8\% insignificant, 14\% minor, 23\% moderate)
\end{itemize} 
Accountability \vspace{0mm}
\begin{itemize}
    \item \textit{Extra constraints} has long-term impacts (53\% major, 29\% catastrophic)
    \item \textit{Lack of ethical knowledge} has short-term impacts (6\% insignificant, 16\% minor, 18\% moderate)
    \item \textit{Lack of technical understanding} has short-term impacts (1\% insignificant, 18\% minor, 21\% moderate)
\end{itemize} 
Fairness \vspace{0mm}
\begin{itemize}
    \item \textit{Extra constraints} has long-term impacts (49\% major, 33\% catastrophic)
    \item \textit{Interpret principles differently} has short-term impacts (3\% insignificant, 17\% minor, 24\% moderate)
\end{itemize} 
Autonomy \vspace{0mm}
\begin{itemize}
    \item \textit{Conflict in practice} has long-term impacts (59\% major, 20\% catastrophic)
    \item \textit{Lack of ethical knowledge} has short-term impacts (6\% insignificant, 13\% minor, 24\% moderate)
\end{itemize} 
Explainability \vspace{0mm}
\begin{itemize}
    \item \textit{Conflict in practice } has long-term impacts (55\% major, 24\% catastrophic)
    \item \textit{Lack of ethical knowledge} has short-term impacts (2\% insignificant, 12\% minor, 27\% moderate)
\end{itemize} 
\\
\hline
RQ3: How these challenges and principles are differently perceived by practitioners and lawmakers? &
\begin{itemize}
    \item The perceptions of practitioners and lawmakers regarding AI ethics principles are strongly correlated (rs=0.819, p=0.000)
    \item The perceptions of practitioners and lawmakers regarding AI ethics challenges are moderately correlated (rs=0.628, p=0.012)
\end{itemize} \\
\hline
\end{tabular}%
\end{table*}

\subsection{Research implications}
\begin{itemize}
    \item We found that most survey respondents agreed to consider the reported principles of AI ethics; however, \textit{transparency}, \textit{accountability} and \textit{privacy} are identified as the most common principles. The study findings complement the existing literature by revealing the most critical principles and call for future research to define the best solutions for scaling the highly significant principles in practice \cite{morley2020initial}.
    \item Regarding the challenges of AI ethics, the survey results confirm the findings of our recent SLR study \cite{AR13}, and determine \textit{lack of ethical knowledge}, \textit{no legal frameworks} and \textit{lacking monitoring bodies} as the high-ranked barriers. The identified challenges are core focus areas that need further research to explore the root causes and best practices to mitigate them.
    \item The study findings indicate that \textit{conflict in practice} is the most severe challenge of AI ethics principles. It opens the door for action-guiding future research - AI ethics principles must be contextualized to balance the conflicts \cite{whittlestone2019role}. The principles need to structure as standards, regulations and codes to resolve the conflicting tensions \cite{whittlestone2019role}.
\end{itemize}

Overall, the study findings complement the emerging research on AI ethics, particularly recognizing the perceptions of two different types of populations (practitioners and lawmakers). Researchers can quickly look up the study results and develop new hypotheses to streamline the mentioned gaps, e.g., solutions to scale the identified principles in practices, explore the causes and mitigation practices of reported challenges, and tailor the existing principles to fit in specific scenarios.
\subsection{Practical implications}
\begin{itemize}
    \item The study findings provide an overview of AI ethics principles, challenges, and practitioners can consider the overall understanding of study findings for defining ethically mature AI processes.
    \item Manifesting AI ethics principles in practice is hard because of various challenging factors. However, we measured the erroneous impacts of these challenges - revealing the most severe barriers practitioners need to tackle before embarking on ethics in AI. 
    \item In general, the study results can facilitate practitioners to get an overview and analyze the extent to which the reported principles and challenges can be leveraged to support AI ethics in the industrial setting.
 \end{itemize}
   
AI ethics in practice is still a widely unexplored research area. We invite researchers from academia and practitioners from industry to jointly contribute by sharing their experiences and to present potential solutions for AI ethics problems. This effort will bridge the gap between academia and industrial practices.

\section{Threats to validity}
\label{threats to validity}
Various threats could affect the validity of the study findings. We followed the guidelines presented in \cite{easterbrook2008selecting} and categorized the potential threats across the following four different types:
\subsection{Internal validity}
Internal validity refers to particular factors that impact methodological rigor. The first internal validity threat in this study is the understandability of the survey instrument. The survey participants might have a different understanding of the survey content; however, the survey instrument was piloted based on the expert's opinions to improve the readability and understandability of the questions (see Section \ref{SettingtheStage}). Moreover, the domain expertise of survey participants could be a potential internal validity threat. We tried to mitigate this threat by exploring various social media networks and used personal links to approach the most suitable candidates. Furthermore, we explicitly mentioned the characteristics of prospective participants in the survey information sheet \cite{replication}. The interpersonal bias in the data collection and analysis process could threaten the internal validity of study findings. However, the survey data is collected, analyzed, organized and reported based on the final consensus made by all the authors (see Section \ref{SettingtheStage} and Section \ref{sec:Results}). 
\subsection{Construct validity}
Construct validity is the extent to which the study constructs are well-interpreted and defended. In this study, AI ethics principles and challenges are the core constructs. The reliability and authenticity of the selected data sources (platforms) is a possible construct validity threat. This threat has been alleviated by searching social media and professional research networks to identify the most relevant groups or individuals. We thoroughly read the group discussions to ensure that the group members mainly discussed AI ethics issues. Similarly, we explored the profile details and interests of the targeted individuals. 
\subsection{External validity}
External validity is the extent to which the study findings based on a particular data sample could be broadly generalized to other contexts. The survey sample size might not be representative to provide a concrete foundation for generalizing the study findings. However, we received 99 valid responses from 20 countries across 5 continents, having a diverse range of experiences, working in various domains on distinct roles in different size organizations (see Figure \ref{Fig:Demographics}). We concede that the study findings could not be generalized at a large scale or consider the identified principles and challenges for all types of AI systems. However, considering the demographic details of the survey respondents (see Figure \ref{Fig:Demographics}), we believe that the study results can support the overall generalizability to some extent.   
\subsection{Conclusion validity}
Conclusion validity is the extent to which certain factors affect valid conclusions in empirical research. To lessen this threat, the first two authors mainly participated in the data collection process; however, the next authors participated in the consent meetings to share feedback and review the survey activities (see Section \ref{SettingtheStage}). Similarly, the third author conducted the data analysis, and the final results were presented based on the feedback shared by all the authors (see Section \ref{sec:Results}). Finally, all the authors were invited to participate in the brainstorming sessions to discuss the core findings and draw concrete conclusions. 

\section{Related Work}
We review the most relevant existing work classified into  two categories, focused on i) AI ethics principles and guidelines, and ii) AI ethics frameworks. A conclusive summary at the end position the scope and contributions of the proposed study.

\subsection{AI ethics principles and guidelines }
Lu et al. \cite{lu2022software} interviewed 21 practitioners and verified that the existing AI ethics principles are broad and do not provide tangible guidance to develop ethically aligned AI systems. Their study findings uncover the fact that AI ethics practices are often considered ad hoc and ignored for continuous learning. Based on the interview findings, Lu et al. \cite{lu2022software} proposed a list of patterns and processes which can be embedded as product features to design a responsible AI system. The proposed design patterns are mainly used to support the core AI ethics principles mentioned by the interview participants. Similarly, Lu et al. \cite{lu2022towards} also conducted an SLR study and defined a software engineering roadmap to develop AI systems. The proposed roadmap covers the entire process life cycle focusing on responsible AI governance, defines process-oriented practices, and presents architectural patterns, styles and methods to build responsible AI systems by design.

Vakkuri et al. \cite{vakkuri2022software} conducted an industrial survey with 249 practitioners to understand and verify the mentioned research gap based on the EU AI ethics guidelines \cite{AR2}. The survey results highlight that most of the companies ignored considering the societal and environmental requirements for developing AI systems. Moreover, the surveyed participants largely considered the product customers as the only stakeholders of AI ethics perspectives; however, it is more narrow in the AI domain, covering customers, regulatory bodies, practitioners, and society. Consequently, the focus should be on multiple AI ethics principles, e.g., \textit{accountability}, \textit{responsibility}, and \textit{transparency}.

Ibanez and Olmeda \cite{ibanez2021operationalising} conducted semi-structured interviews with 22 practitioners and two focus groups to know how software development organizations address ethical concerns in AI systems. The interview findings raised various issues related to AI ethics principles and practice including \textit{governance}, \textit{accountability}, \textit{privacy}, \textit{fairness}, and \textit{explainability}. Moreover, the interview participants provide some suggestions to operationalize AI ethics, e.g., promoting domain focus standardization, embracing data-driven organizational culture, presenting a particular code of ethics and fostering AI ethics awareness. In conclusion, Ibanez and Olmeda \cite{ibanez2021operationalising} called for a set of actions to distinguish project stakeholders, develop a socio-technical project team, and regularly evaluate the AI projects practices, processes and policies.

\subsection{AI ethics frameworks}
Vakkuri et al. \cite{vakkuri2021eccola} developed the ECCOLA framework to provide a tool for implementing ethics in AI. ECCOLA aims to assist practitioners, and AI-specific software development organizations in adopting ethically aligned development processes. The proposed framework supports iterative development and consists of a deck of cards (modules) that could be tailored to a specific context. The card mainly defines various themes of AI ethics, which were identified in the existing AI ethics guidelines. The ECCOLA framework is evaluated in both the academia and industrial domain to understand its real-world implications and limitations. 

Floridi et al. \cite{floridi2018ai4people} proposed the AI4People framework comprising five principles and twenty recommendations to lay the foundation for “Good AI Society”. The available sets of AI ethics principles are comparatively synthesized to understand the commonalities and significant differences. The comparison findings reveal four AI ethics principles (\textit{beneficence}, \textit{non‑maleficence}, \textit{autonomy}, \textit{justice}) with a new additional principle  (\textit{explicability}) to structure the AI4People framework. Finally, twenty action points were devised to scale the mentioned principles in practice. The overall aim of the proposed framework is to move the dialogue forward from theoretical principles to in-action policies. Such policies shield human autonomy, increase social empowerment, and decrease inequality.

Leikas et al. \cite{leikas2019ethical} presented a framework that focuses on ethics by design in decision-making systems. The current design approaches, practices, theories and concepts of autonomous intelligent systems are reviewed to structure the proposed ethical framework. The framework could use to recommend a set of AI ethics principles and practices for a specific scenario. The framework captured the human-centric details of a particular case study and used the details to identify the ethical requirements of concerned stakeholders and transfer them to the design goals. Leikas et al. \cite{leikas2019ethical} called for future studies to evaluate the real-world significance of the proposed framework in industrial scenarios.

\subsection{Conclusive summary}
The reported studies \cite{lu2022software}\cite{lu2022towards}\cite{vakkuri2022software}\cite{ibanez2021operationalising} are grounded on empirical findings and fine-granular analysis of extant AI ethics principles and guidelines. To complement empiricism in exploring AI ethics principles and challenges, this study explicitly analyzed and discussed the principles based on the perceptions of two different types of populations (practitioners and lawmakers). Studies \cite{vakkuri2021eccola}\cite{floridi2018ai4people}\cite{leikas2019ethical} are conducted to  design various frameworks to operationalize the AI ethics principles; however, no research has yet been done to streamline the plethora of challenges in adopting the widely defined principles and frameworks. Our study preliminarily focused on survey-driven validation of AI ethics principles and challenges by practitioners and lawmakers to complement the body of research comprising the recent industrial studies on AI ethics principles \cite{lu2022software}\cite{lu2022towards}\cite{vakkuri2022software}\cite{ibanez2021operationalising} and frameworks \cite{vakkuri2021eccola}\cite{floridi2018ai4people}\cite{leikas2019ethical}.
\section{Conclusions and future work}
\label{LessonsLearned}
This empirical study explored the perceptions of representative practitioners and lawmakers on AI ethics principles and potential challenges. We outlined the following observations based on the data collected from 99 respondents working in 20 different countries on various roles with diverse working domains:

\underline{\textbf{Emerging Roles}}: Besides practitioners, the role of policy and lawmakers is also important in defining the ethical solutions for AI-based systems. Based on our knowledge, this study is the first effort made to encapsulate the views and opinions of both types of populations.

\underline{\textbf{Confirmatory Findings}}: This study empirically confirms the AI ethics principles and challenging factors discussed in our published SLR study \cite{AR13}. Based on the survey findings, most participants agreed that the identified principles and challenges should take into consideration for defining ethics in AI.

\underline{\textbf{Adherence to AI principles and challenges}}: The most common principles \textit{(e.g., transparency, privacy, accountability)} and challenges \textit{(e.g., lack of ethical knowledge, no legal frameworks, lacking monitoring bodies)} must be carefully realized in AI ethics. Companies must consider the mentioned common principles and challenges to define ethically aligned design methods and frameworks in practice.

\underline{\textbf{Risk-aware AI ethics}}: The challenging factors have mainly long-term severity impacts across the AI ethics principles. It opens a new research call to identify the causes of the most severe challenging factors and propose solutions for minimizing or mitigating their impacts.

\underline{\textbf{Practitioners and lawmakers perceptions}}: The identified principles and challenges are statically analyzed to understand the significant differences between practitioners’ and lawmakers' perceptions. We noticed that the opinions of both types of populations are positively and significantly correlated. In the long term, these findings could use to develop lawful (complying with applicable laws) and robust (technically and socially) AI ethics solutions (adhering to ethical principles)\cite{AR2}.

\underline{\textbf{Future research}}: Our final catalogue (see Figure \ref{Fig:SurveyFindings}) of principles and challenging factors can be used as a guideline for defining ethics in the AI domain. Moreover, the catalogue is a starting point for further research on AI ethics. It is essential to mention that the identified principles and challenging factors only reflect the perceptions of 99 practitioners and lawmakers in 20 countries. More deep and comprehensive empirical investigation with wider groups of practitioners to discuss the causes and solutions of the identified challenges would be useful to generalize the study findings at large scale. This, together with proposing a robust solution (AI ethics maturity model) for integrating ethical aspects in AI design and process flow, will be part of our future work.

\bibliographystyle{IEEEtran}
\bibliography{References.bib}

\end{document}